\documentclass[12pt]{article}

\author{Vladimir A. Petrov \footnote{e-mail:Vladimir.Petrov@ihep.ru}}

\title{Can the Odderon be "Supercritical"?}
\date{}

\begin{document}

\maketitle
A. A. Logunov Institute for High Energy Physics, NRC KI,
Protvino, RF

\begin{abstract}
Based on the postulate that a $ C $-odd exchange leads to repulsion in the particle-particle system (say, $ pp $) and to attraction in the antiparticle-particle system (say, $ \bar{p}p $), we show that the Odderon intercept cannot be greater than unity.
\end{abstract}

\section*{Introduction}	
Over the long history of studying the "Odderon" a whole spectrum of its intercept values has been given (if by Odderon we mean the $ C $ -odd Regge trajectory, $\alpha_{\mathcal{O}}(t)$): from $ \alpha_{\mathcal{O}}(0)< 1 $ (through $ \alpha_{\mathcal{O}}(0)=1 $) to $ \alpha_{\mathcal{O}}(0) > 1 .$ 
All these results, however, are unsatisfactory because of their RG non-invariance\footnote{The widespread misunderstanding that the Regge intercept of a Regge trajectory in massless QCD is \textit{independent} of the QCD coupling constant $ \alpha_{s}^{QCD} $ has been the cause of numerous failed attempts to calculate the Pomeron and Odderon intercepts as functions of $ \alpha_{s}^{QCD} $.}. 

While waiting for a satisfactory solution to the value of the Odderon intercept in QCD, we set ourselves a more modest goal:  to analyze the question of the sign of the quantity 
\[\alpha_{\mathcal{O}}(0)-1\]
based only on the above-mentioned correspondence between the character of strong forces and C-parity.

 Keeping in mind the accompanying "unitarization" (eikonal representation) and for reference we remind the standard expression for the Odderon as a "Born amplitude" $ \mathcal{O}(s,t) $ defined by the negative signature Regge pole at $ j=\alpha_{\mathcal{O}}(t) $\footnote{We consider for definiteness $ pp $-scattering.}  (see e.g. Ref. \cite{Van})    
\begin{equation}
\mathcal{O}(s,t) = \pi\frac{\alpha_{\mathcal{O}}^{'}(t)(\alpha_{\mathcal{O}}(t)+1/2)}{sin(\pi\alpha_{\mathcal{O}}(t))}[1-exp(-i\pi \alpha_{\mathcal{O}}(t)]\Gamma_{\alpha_{\mathcal{O}}(t)}^{2}(t)P_{\alpha_{\mathcal{O}}(t)}(-z_{t})=
\end{equation}
\[= \pi\alpha_{\mathcal{O}}^{'}(t)(\alpha_{\mathcal{O}}(t)+1/2)[i+ tg\frac{\pi\alpha_{\mathcal{O}}(t)}{2}]\Gamma_{\alpha_{\mathcal{O}}(t)}^{2}(t)P_{\alpha_{\mathcal{O}}(t)}(-z_{t})\]
where $ \Gamma_{\alpha_{\mathcal{O}}(t)}(t) $ is the spin $ J $ meson vertex $ \Gamma_{J}(t) $
at $ J=\alpha_{\mathcal{O}}(t) $ and $z_{t}=1-2s/(4m^{2}-t)$.
 When moving to the $ t $-channel, $ \Re\alpha_{\mathcal{O}}(t)\rightarrow J = 1,3,... $, and the propagator of the C-odd meson of (generally complex) mass $ m_{J} $ is reproduced:
\begin{equation}
\mathcal{O}(s,t)\sim \frac{\Gamma_{J}^{2}(t)}{m^{2}_{J} - t} P_{J} (z_{t}).
\end{equation}
The propagator stems from $ tg(\pi\alpha_{\mathcal{O}}(t)/2). $
 There are no poles (physical states) in the cross-channel of amplitudes of the "alien"  (in this case positive, with even $J  $) signature.
\section*{Odderon accompanies Pomeron: \\ add or subtract?}

Assuming that at sufficiently high energies the contribution of secondary meson trajectories is negligibly small, we will use, as the Born term, along with the Odderon contribution, the Pomeron contribution, which is the leading one according to the unitarity condition, viz.

\begin{equation}
\mathcal{P}(s,t) = -\pi\alpha_{\mathcal{P}}^{'}(t)(\alpha_{\mathcal{P}}(t)+1/2)\frac{1+exp(-i\pi \alpha_{\mathcal{P}}(t)}{sin(\pi\alpha_{\mathcal{P}}(t))}\Gamma_{\alpha_{\mathcal{P}}(t)}^{2}(t)P_{\alpha_{\mathcal{P}}(t)}(-z_{t})
\end{equation}
Eikonal functions are related to Born terms as
\begin{equation}
\delta_{\mathcal{P(O)}}(s,b)= \frac{1}{16\pi s}\int dt J_{0}(b\sqrt{-t})\mathcal{P(O)}(s,t)
\end{equation}
Following the postulate formulated in the Abstract, we must assume that
\begin{equation}
\delta_{\bar{p}p}= \delta_{\mathcal{P}}(s,b)+\delta_{\mathcal{O}}(s,b)
\end{equation}
and
\begin{equation}
\delta_{pp}= \delta_{\mathcal{P}}(s,b)-\delta_{\mathcal{O}}(s,b)
\end{equation}
\section*{Potentials}
To make the choice of signs in (5) and (6) more visual, we use the connection between eikonal phases and interaction potentials. The latter are related to the Born amplitudes
as
\begin{equation}
V_{\mathcal{P/O}}= - \frac{1}{2s}\int \frac{d^{3}q}{(2\pi)^{3}} [\mathcal{P/O}(s,t)]\exp{ {\Large i} \textbf{r}\textbf{q}}, \textbf{q}^{2} = -t.
\end{equation}

The Pomeron term leads to the potential
\begin{equation}
V_{\mathcal{P}}\approx -g^{2}_{P}(i+tan\frac{\pi\Delta_{\mathcal{P}}}{2})exp(-r^{2}/R^{2}_{\mathcal{P}})
\end{equation}
while the Odderon one to
\begin{equation}
V_{\mathcal{O}}\approx -g^{2}_{O}(i-cot\frac{\pi\Delta_{\mathcal{O}}}{2})exp(-r^{2}/R^{2}_{\mathcal{O}}).
\end{equation}

Here 
\[\Delta_{\mathcal{P(O)}}=\alpha_{\mathcal{P,O}}(0)-1\]
A simple analysis of equations (7) and (8) shows that the postulate from the Abstract is satisfied only under the condition that
\[\Delta_{\mathcal{P}} > 0; \Delta_{\mathcal{O}} < 0 .\]
Comparison of the imaginary parts of the complex potentials shows that central absorption is stronger in the $ \bar{p}p $ channel than in the $ pp $ channel:
\begin{equation}
\Im V_{\bar{p}p}\sim g^{2}_{P} + g^{2}_{O},
\end{equation}
\begin{equation}
\Im V_{pp} \sim  g^{2}_{P} - g^{2}_{O}.
\end{equation}

\section*{"Critical" case: $\Delta_{\mathcal{O}}=0$.}

The case $\alpha_{\mathcal{O}}(t)= 1+ \alpha^{'}_{\mathcal{O}}(0)t +...  (\Delta_{\mathcal{O}}=0) $  needs a special analysis.
Here we get
\begin{equation}
\mathcal{O}(s,t)\mid_{t\rightarrow 0} \approx  const\: \Gamma_{\alpha_{\mathcal{O}}(t)}^{2}(t)[\frac{1}{-t} + i\frac{\pi}{2\alpha^{'}_{\mathcal{O}}(0)}]s.
\end{equation}
The first term in the square brackets would mean the presence in the cross channel $ \bar{p}p $ of a massless vector meson ( "strong photon")
 which is inadmissible.
To avoid this we have to assume that $\Gamma_{\alpha_{\mathcal{O}}(t)}^{2}(t)\sim (-t)^{N}  $ with $ N $ integer and $ N \geq 1 $ as follows from analyticity in $ t $.
This condition must be taken into account in models when using $ \mathcal{O}(s,t) $ as part of the eikonal in phenomenological data processing.

It must be said that taking into account decoupling factors makes the expressions for the eikonal in the space of impact parameters, $\delta_{\mathcal{O}}(s,b)$, rather cumbersome.

In addition, one should be aware that "forward decoupling" of the  Odderon in the eikonal does not mean the same decoupling in full amplitude.

Let us add that we do not discuss the formally possible option with a fixed pole $  \sim g(t)/(J-1) $ (i.e. $\alpha_{\mathcal{O}}(t)=1, \forall t $ ) as it is physically meaningless.
\section*{Discussion and Conclusions}

Thus, if we assume, as is customary, that C-even exchanges correspond to attractive forces, while C-odd exchanges lead to repulsive forces between hadrons $ hh $ and attractive forces between $ \bar{h}h$, then the Odderon intercept cannot surpass $ 1 $:

\begin{equation}
\alpha_{\mathcal{O}}(t)\leq 1.
\end{equation} In some jargon one could say that "the Odderon cannot be supercritical."

We also have shown that from the same premise it follows "
forward decoupling" of the "critical" Odderon ($ \alpha_{\mathcal{O}}(0)= 1 $) from the Born amplitude underlying the eikonal phase.

Of course, this does not mean that the unsatisfactory results mentioned in the Introduction  are correct, despite the fact that they give
$ \alpha_{\mathcal{O}}(0)\leq 1 $.

 \section*{Acknowledgements}
 I am thankful to Nikolay P. Tkachenko and Anatolii K. Likhoded for discussions.

\end{document}